\documentclass[
  reprint,
  superscriptaddress,
  amsmath, amssymb,
  aps,
  pra,
  floatfix,
  nofootinbib, 
]{revtex4-2}

\usepackage[letterpaper, left=18mm, right=18mm, top=25mm, bottom=25mm, columnsep=22pt]{geometry}

\usepackage[utf8]{inputenc}
\usepackage[T1]{fontenc}
\usepackage[english]{babel}
\usepackage{lmodern}
\usepackage[colorlinks=true,citecolor=blue,linkcolor=blue,urlcolor=blue]{hyperref}
\usepackage{graphicx}
\usepackage{float}
\usepackage[caption=false]{subfig}
\usepackage{color,xcolor}
\usepackage{tikz}
\usepackage{braket}
\usepackage{bm}
\usepackage{comment}
\usepackage{physics}
\usepackage{mathtools}
\usepackage{enumitem}
\usepackage{appendix}
\usepackage{multirow}
\usepackage{booktabs}
\usepackage{glossaries}
\usepackage{amssymb}
\usepackage{amsmath}
\usepackage{amsfonts}
\usepackage{quantikz}
\usepackage{cancel}
\usepackage{dsfont}
\usepackage{soul}
\usepackage{xfrac}
\usepackage{times}

\definecolor{gatecolor}{HTML}{6C3C84}
\definecolor{pulsecolor}{HTML}{6C3C84}



\newacronym{qml}{QML}{quantum machine learning}
\newacronym{pqml}{PQML}{pulsed quantum machine learning}
\newacronym{ai}{AI}{artificial intelligence}
\newacronym{gpu}{GPU}{Graphics Processing Unit}
\newacronym{ml}{ML}{machine learning}
\newacronym{ann}{ANN}{Artificial Neural Networks}
\newacronym{nn}{NN}{neural networks}
\newacronym{qnn}{QNN}{quantum neural network}
\newacronym{qoc}{QOC}{Quantum optimal control}
\newacronym{eqk}{EQK}{embedding quantum kernel}
\newacronym{nisq}{NISQ}{noisy intermediate-scale quantum}
\newacronym{vqc}{VQC}{variational quantum circuits}
\newacronym{qaoa}{QAOA}{Quantum Approximate Optimization Algorithm}
\newacronym{vqe}{VQE}{Variational Quantum Eigensolver}
\newacronym{cpb}{CPB}{cooper pair box}
\newacronym{cr}{CR}{cross-resonance}
\newacronym{vz}{VZ}{virtual Z}
\newacronym{rwa}{RWA}{Rotating Wave Approximation}
\newacronym{qho}{QHO}{Quantum Harmonic Oscillator}

\newcommand{\identity}{\mathbb{\hat I}}

\newcommand{\ie}{{\it{i.e.~}}}

\newcommand{\sigmax}{\hat\sigma^x}
\newcommand{\sigmay}{\hat\sigma^y}
\newcommand{\sigmaz}{\hat\sigma^z}

\begin{document}

\title{Pulsed learning for quantum data re-uploading models}

\author{Ignacio B. Acedo}
\email[Corresponding author: ]{\qquad ibenito@quantum-mads.com}
\affiliation{Quantum Mads, Calle Larrauri 1, Edificio A, piso 3, puerta 28, 48160 Derio, Spain}
\affiliation{Departamento de Física, Universidad Carlos III de Madrid, Avda. de la Universidad 30, 28911 Leganés, Spain}

\author{Pablo Rodriguez-Grasa}
\affiliation{Department of Physical Chemistry, University of the Basque Country UPV/EHU, Apartado 644, 48080 Bilbao, Spain}
\affiliation{EHU Quantum Center, University of the Basque Country UPV/EHU, Apartado 644, 48080 Bilbao, Spain}
\affiliation{TECNALIA, Basque Research and Technology Alliance (BRTA), 48160 Derio, Spain}

\author{Pablo Garcia-Azorin}
\affiliation{Department of Physical Chemistry, University of the Basque Country UPV/EHU, Apartado 644, 48080 Bilbao, Spain}
\affiliation{EHU Quantum Center, University of the Basque Country UPV/EHU, Apartado 644, 48080 Bilbao, Spain}
\date{\today}

\author{Javier Gonzalez-Conde}
\affiliation{Quantum Mads, Calle Larrauri 1, Edificio A, piso 3, puerta 28, 48160  Derio, Spain}
\affiliation{Department of Physical Chemistry, University of the Basque Country UPV/EHU, Apartado 644, 48080 Bilbao, Spain}

\begin{abstract}
While Quantum Machine Learning (QML) holds great potential, its practical realization on Noisy Intermediate-Scale Quantum (NISQ) hardware has been hindered by the limitations of variational quantum circuits (VQCs). Recent evidence suggests that VQCs suffer from severe trainability and noise-related issues, leading to growing skepticism about their long-term viability. However, the possibility of implementing learning models directly at the pulse-control level remains comparatively unexplored and could offer a promising alternative. In this work, we  formulate a pulse-based variant of data re-uploading, embedding trainable parameters directly into the native system’s dynamics. We benchmark our approach on a simulated superconducting transmon processor with realistic noise profiles. The pulse-based model consistently outperforms its gate-based counterpart, exhibiting higher test accuracy and improved generalization under equivalent noise conditions. Moreover, by systematically increasing noise strength, we show that the pulse-based model retains useful classification accuracy at depolarizing probabilities where its gate-based counterpart has already degraded to near-random performance. These results position pulse-native architectures as a viable and hardware-aligned path toward practical QML on current devices.
\end{abstract}

\maketitle

\section{Introduction}\label{sec:introduction}

In recent years, \gls*{qml} has gained significant attention due to its promise to merge quantum computational power with the transformative impact of machine learning \cite{RASHID2024100277, gill2024quantumcomputingvisionchallenges, qml, challenges_QML, qml_tutorial}. While early proposals of fault-tolerant algorithms suggested exponential advantages over classical methods \cite{QSVM, quantum_PCA}, it was later recognized that the underlying assumptions of these schemes make such advantages less evident \cite{dequant}. Consequently, the community’s focus shifted toward realizations compatible with existing quantum devices, where a proof-of-concept demonstration of quantum advantage was also reported \cite{rigorous_speedup}. In this sense, current devices still belong to the \gls*{nisq} era~\cite{Preskill2018quantumcomputingin}, characterized by limited qubit counts, short coherence times, and relatively high error rates.

Within this landscape, \gls*{vqc} have emerged as the main paradigm for leveraging \gls*{nisq} hardware \cite{Benedetti_2019, vqas}. Among the techniques developed to enhance their performance, data re-uploading has become particularly prominent \cite{dataReuploading}. By repeatedly embedding the input data at multiple stages of the circuit and alternating these embeddings with trainable data-processing operations, this strategy enriches the model’s expressive power while maintaining shallow circuit depths. Thus, the combination of hardware efficiency and improved representational capabilities made the data re-uploading model an appealing approach for near-term quantum learning tasks \cite{schuld_FT}. However, several challenges limit the practical usefulness of conventional variational models in realistic settings. A first issue arises from noise, which restricts their expressive capacity and the complexity of the solutions they can reliably achieve \cite{expressive_measure}. It has also been observed that standard QML architectures can memorize randomly labeled data, highlighting the need for careful model design when generalization is expected \cite{Gil-Fuster2024}. In addition, recent studies indicate that certain gate-based QML proposals may offer no inherent advantage over classical methods once their structure and data-encoding strategies are examined \cite{Bowles2024BetterTC}. These considerations have motivated a broader rethinking of how quantum learning models should be constructed, encouraging the exploration of new approaches that move beyond conventional gate-based parametrizations.

On the other hand, learning models implemented directly at the pulse-control level, the native language of the hardware, remain relatively unexplored, yet they offer a promising alternative to standard gate-based approaches~\cite{meitei2021gatefreestatepreparationfast}. Working with pulses allows control signals to be tailored to the actual dynamics of the qubits, avoiding the inefficiencies and errors introduced by compiling abstract gate sequences~\cite{quantumGuide, Alexander_2020}. This leads to more accurate state manipulations and higher-fidelity operations, and it enables the design of controls that naturally account for device-specific features such as frequency detuning, anharmonicity, or qubit connectivity~\cite{10.3389/fphy.2022.900099} 

Pulse-level parametrization also provides additional degrees of freedom for encoding information and for optimizing the model, which can improve the structure of the optimization landscape and potentially reduce the severity of barren plateaus~\cite{barrenplateauspulses}. Finally, by compressing multi-gate sequences into shorter pulse schedules, this approach reduces circuit depth and execution time~\cite{PhysRevLett.129.060501}. Overall, pulse-level control offers a flexible and hardware-aware framework for building problem-specific ansätze that make more efficient use of current quantum hardware. 

So far, the development of hardware-native quantum algorithms has motivated a gradual shift from gate-level formulations toward pulse-level optimization tailored to specific platforms. Early contributions such as ctrl-VQE \cite{pulsed_ctrl_VQE} demonstrated that pulse attributes (amplitude, frequency, and phase) can be treated as variational parameters, avoiding explicit gate decompositions. Subsequent work has explored pulse-parameterized training strategies \cite{PhysRevApplied.23.024036}, cross-resonance implementations \cite{PhysRevResearch.5.033159}, pulse-level variants of QAOA \cite{pulsed_optimization}, and pulse-efficient transpilation techniques \cite{Meirom_2023, Melo2023pulseefficient}. More recently, studies have analyzed pulse design spaces in terms of expressivity, entanglement generation, and trainability \cite{liang2022variationalquantumpulselearning, liang2024advantagesparameterizedquantumpulses, tao2025designexpressivetrainablepulsebased}.

Despite this progress, two limitations remain common. First, many practical approaches still inherit a gate-defined structure, where pulses mainly reparameterize pre-existing circuit templates. Second, more Hamiltonian-inspired formulations like \cite{tao2024unleashingexpressivepowerpulsebased}, while insightful, are often developed under assumptions that make direct deployment on realistic multi-qubit hardware less straightforward. As a result, there is still a gap between theoretical pulse-based advantages and scalable, hardware-compliant QML implementations.

This motivates the need for pulse-native learning frameworks that jointly address control design, trainability, and hardware constraints, while remaining compatible with realistic near-term devices.

In this work, we extend the data re-uploading paradigm to the pulse-control level. Our framework provides a general methodology for translating any gate-based variational QML architecture into a pulse-level implementation: parameterized gates are replaced by native control pulses whose physical attributes (amplitude, phase, and detuning) serve as trainable parameters, embedding learning directly into the driven system's dynamics. Importantly, this is not a reparameterization of a fixed gate template; it redesigns the trainable layer at the hardware level while keeping the encoding strategy modular.

As a proof of concept, we instantiate this framework on a data re-uploading QNN for superconducting transmon qubits and benchmark it under realistic noise extracted from the IBM Brisbane device. Our main findings are: (i) the pulse-based model generalizes better, achieving comparable test accuracy with fewer layers and exhibiting a delayed onset of overfitting; (ii) it is more noise-resilient, sustaining meaningful accuracy at depolarizing strengths where the gate-based model collapses to chance level; and (iii) these advantages hold consistently across four distinct binary classification tasks (MNIST digits, Iris, Corners, and Helix).

The manuscript is organized as follows. Section~\ref{sec:preliminaries} introduces the theoretical background underlying our pulse-based models, including the fundamentals of QOC for superconducting platforms and the principles of data re-uploading in circuit-based QML. Section~\ref{sec:proposed_model} presents our pulse-based extensions and their application to the aforementioned QNN architecture. Section~\ref{sec:results} details the simulation setup and results, emphasizing the practical advantages of our approach under realistic noise. Finally, Section~\ref{sec:conclusions} summarizes our findings and outlines promising directions for future research on pulse-level QML.

\section{Preliminaries}\label{sec:preliminaries}
Our approach combines insights from two complementary domains. On the one hand, quantum optimal control provides the tools to describe and manipulate the continuous-time dynamics of qubits at the pulse level, allowing us to design parametrizations that are closer to the hardware and more expressive than gate abstractions. In this work we focus on a proposal based on superconducting transmon qubits~\cite{Transmon_original}. On the other hand, quantum machine learning introduces algorithmic strategies such as data re-uploading, which enable efficient feature encoding and trainable variational models. Bringing these perspectives together establishes a foundation for constructing pulse-based ansätze that are both physically grounded and well-suited for realizing effective algorithms on \gls*{nisq} devices

\begin{figure}[t!]
    \centering
    \includegraphics[width=0.85\columnwidth]{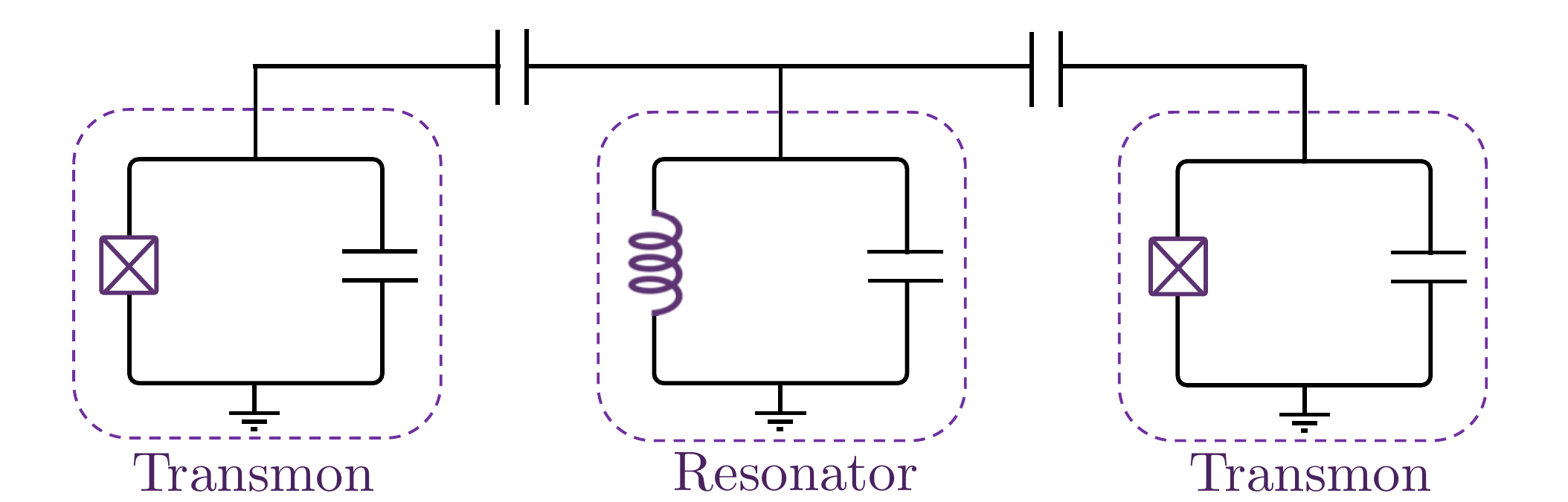}
    \caption{Capacitive coupling of two transmon qubits via a coupler in form of a linear resonator. Each transmon is made with a Josephson junction and a capacitor. The transmon qubits interact through an intermediate resonator, which is composed of an inductance and a capacitor.}
    \label{fig:coupler}
\end{figure}

\subsection{Quantum Optimal Control}\label{subsec:qoc}

Quantum optimal control is a field that focuses on manipulating and optimizing the dynamics of quantum systems through external controls, typically in the form of electromagnetic fields, to achieve desired quantum states or operations~\cite{quantum_control_book}. By exploiting the principles of quantum mechanics, QOC techniques allow for precise adjustments to qubit behavior, enhancing the fidelity and performance of quantum computations.

Pulse programming, a crucial aspect of \gls*{qoc}, represents the most direct and fundamental way to interact with quantum hardware, essentially ``\textit{communicating}'' with quantum computers in its native language. Unlike gate-level programming, which relies on abstracted quantum operations carefully calibrated regularly and built from limited predefined blocks, pulse programming gives precise control over the electromagnetic signals that manipulate qubit dynamics at a fundamental level. This low-level access provides a deeper understanding of quantum hardware, enabling researchers to fine-tune quantum gates, reduce latency and error rates, increase the accuracy of certain processes, and design custom operations that surpass the limitations of conventional fixed set of allowed
operations~\cite{Alexander_2020}.

In this section, we introduce the key concepts of pulse programming on superconducting transmon qubits, laying the groundwork for constructing a general pulsed-QML model.

\subsubsection{Transmon qubits}

{Up to the moment of realization of this work, superconducting quantum computers represent one of the most advanced quantum computing technologies~\cite{willow, superconducting_google, superconducting_ibm}. 
In this platform, the information is encoded in the quantum degrees of freedom of nanofabricated superconducting circuit elements. At the core of this technology are the so-called \emph{transmon qubits}, anharmonic oscillators built out of a Josephson tunnel junction connected to a large shunting capacitor, creating a coherent and controllable system capable of encoding a qubit of information~\cite{Transmon_original}. A simple illustrative figure representing the basic circuit just described is displayed in Fig.~\ref{fig:coupler} and its theoretical modelization can be found in Appendix~\ref{sec:Appendix QOC}.}

The transmon system is manipulated using microwave pulses, which enables precise control over the state of the anharmonic oscillator. By tuning the pulse parameters, the population distribution across its energy levels can be accurately adjusted. After applying certain approximations and transformations~\cite{quantumGuide, CRHamiltonian_errors}, the driven effective Hamiltonian in the interaction picture of a single transmon qubit can be expressed as
\begin{equation}\label{eq:one_pulse_ham}
    \hat H_d =-\frac{\Delta_{\omega}}{2} \sigmaz +  \frac{\Omega s(t)}{2}(\cos \gamma \;\sigmax + \sin\gamma\; \sigmay) ,
\end{equation}
where the applied pulse takes the form  $\Omega s(t) \cos(\omega_d t + \gamma)$. Here,  $\Omega$ denotes the driving pulse amplitude, $s(t)$ the pulse shape, $\omega_d$ is the driving frequency and $\gamma$ denotes the phase. Finally, we define $\Delta_{\omega} = \omega_q - \omega_d$ as the \textit{detuning} between the qubit and the drive frequencies, $\omega_q$ and $\omega_d$ respectively.

\begin{figure}[t!]
    \centering
    \includegraphics[width=0.85\columnwidth]{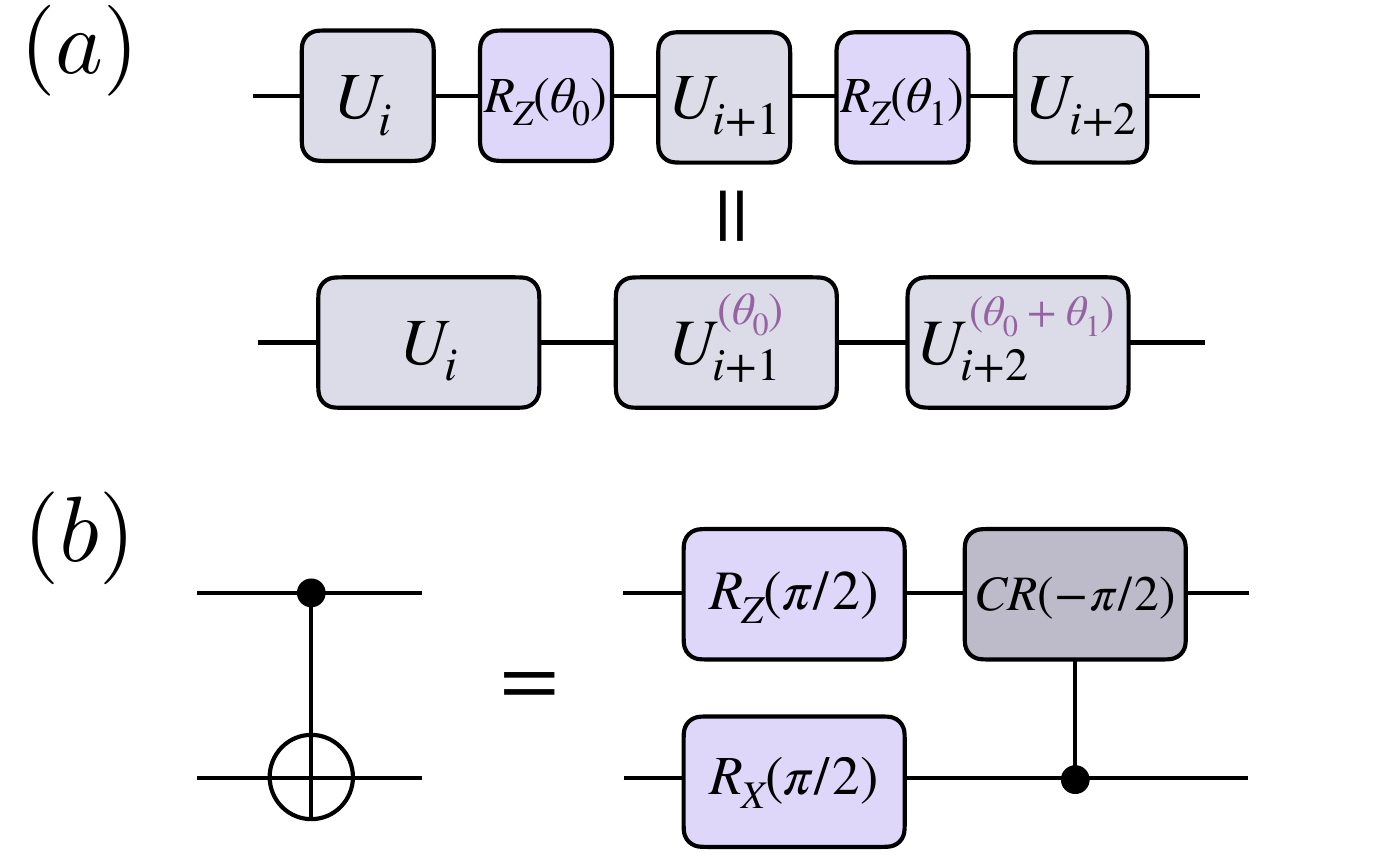}
    \caption{ (a) Virtual Z rotations scheme. $U$ gates represent an arbitrary gate. The upper indices denote the phase offset of the pulse. (b) CNOT gate from the three basic blocks for transmon. Here CR represents a cross resonance gate.
Virtual rotations and CNOT gate transpilation with transmon architecture.}
    \label{fig:transmon_gates}
\end{figure}

From this Hamiltonian, the abstraction of gate operations becomes apparent, as it describes the interaction between the qubit and an external field. By choosing a pulse with a frequency resonant with the qubit frequency, \ie $\Delta_\omega = 0$, one can perform rotations in the $XY$-plane where the axis of rotation is determined by the pulse phase, $\gamma$, and the rotation angle is controlled by the duration and the Rabi frequency of the pulse, $\Omega s(t)$. For example, setting $\gamma = \pi/2$ enables the implementation of $Y$-rotations.

On the other hand, exact rotations around the $Z$ axis can be achieved using a technique known as \gls*{vz} gates~\cite{vitualZgates}. This method enables perfect $Z$-rotations without requiring physical operations. Instead, the rotation is \textit{virtually} implemented by introducing a phase shift to subsequent pulses as illustrated in Fig.~\ref{fig:transmon_gates}\textcolor{blue}{(a)}. Consequently, \gls*{vz} rotations are both cost-free and error-free, as they are executed instantaneously. Additionally, recent work has studied the implementation of this virtual gate for its direct pulsed implementation~\cite{long2025virtualzgatesvirtual}. See Appendix~\ref{sec:Appendix QOC} for a more detailed discussion.

These two operations—$XY$-rotations with frequency-tuned pulses and \gls*{vz} rotations—enable universality in the Hilbert space of a single qubit, as any SU(2) unitary matrix, $ U$, can be decomposed into three rotations around the $Y$ and $Z$ axes, 
\begin{equation}\label{eq:SUrotationDecomposition}
    U = R_Z(\theta_1) R_Y(\theta_2) R_Z(\theta_3).
\end{equation}
These rotation angles are commonly referred to as \textit{Euler angles}.

\subsubsection{Interaction between transmon qubits}

Qubits can be interconnected in various ways to enable interactions. Inspired by the IBM quantum computer implementation, this work adopts capacitive coupling mediated via a linear resonator~\cite{quantumGuide}, eliminating the need for flux tunable coupling elements. A schematic representation of this coupling mechanism is shown in Fig.~\ref{fig:coupler}  and its modelization is discussed in Appendix~\ref{sec:Appendix QOC}. This type of fixed-frequency qubit architecture is well-suited for implementing the well-known two-qubit entangling gate known as \gls*{cr} gate.

The \gls*{cr} gate is a microwave-activated two-qubit entangling operation commonly employed in fixed-frequency, off-resonantly coupled qubits. It functions by applying a pulse to the control qubit at the target qubit’s frequency, inducing an effective $XZ$-rotation~\cite{PhysRevA.101.052308}. Despite the mathematical complexity underlying its derivation, the essential mechanism can be captured by the effective Hamiltonian:

\begin{equation}\label{eq:cr_gate_ham}
\begin{aligned}
\hat H_{\text{CR}} = & -\frac{\Delta_{12}}{2} \sigmaz \otimes \identity \\
+ \frac{\Omega(t)}{2} &\Big[ \cos \gamma \left( \sigmax \otimes \identity + \mu\ \sigmaz \otimes \sigmax + \nu \;\identity \otimes \sigmax \right) \\
\qquad \quad &+ \sin \gamma \left( \sigmay \otimes \identity + \mu \;\sigmaz \otimes \sigmay + \nu \;\identity \otimes \sigmay \right) \Big].
\end{aligned}
\end{equation}
where $\Delta_{12} = \omega_{q,1} - \omega_{q,2}$ is the difference between the two qubit frequencies, the coefficient $\nu$ represents the cross-talk effect, and $\mu$ the CR-interaction term, see Appendix \ref{sec:Appendix QOC} for further details.

The \gls*{cr} gate's simplicity, and reliance on microwave control, make it an attractive option for implementing two-qubit gates in fixed-frequency qubit architectures. However, achieving high-fidelity operations requires careful calibration to mitigate errors such as crosstalk and leakage~\cite{CRHamiltonian_errors}. Recent advances in pulse control techniques have improved gate fidelities and reduced operation times, further enhancing the practicality of the CR gate for near-term quantum computing applications~\cite{PRXQuantum.2.040336}. For the purposes of this work, the \gls*{cr} gate provides a sufficient mechanism to induce entanglement within our pulse-level \gls*{qml} ansätze, enabling the construction of expressive and hardware-aligned quantum models without the need for additional entangling mechanisms.

In this work we are going to consider three fundamental pulse operations to manipulate the superconducting quantum processor: a pulse resonant with the qubit frequency for single-qubit $XY$-rotations (Eq.~\ref{eq:one_pulse_ham}), \gls*{vz} rotations to implement Z-axis single-qubit rotations (Fig.~\ref{fig:transmon_gates}{\color{blue}$(a)$}), and the \gls*{cr} gate to entangle qubits and enable multi-qubit rotations (Eq.~\ref{eq:cr_gate_ham}). Any multi-qubit operation can be transpiled using these three foundational building blocks. For example, the CNOT gate can be constructed in this manner, as illustrated in Fig.~\ref{fig:transmon_gates}{\color{blue}$(b)$}.

\subsection{Data re-uploading }\label{subsec:qml}

A wide variety of parameterized quantum circuits for machine learning tasks have been proposed in recent years, such as the hardware efficient ansatz \cite{Leone2024practicalusefulness}. The focus of this work, however, does not lie in selecting the most effective model. Instead, our goal is to employ a manageable architecture in order to demonstrate how a pulse-level adaptation can yield a model that is more robust to noise and capable of better generalization compared to a gate-level trained model. From this perspective, our approach may also serve as motivation for adapting other models from the literature directly to the hardware level. 

We address the task of binary classification as a representative benchmark.
For this study, we adopt a \gls*{qnn} architecture previously introduced in \cite{pablo_kernel}, which is based on a data re-uploading scheme \cite{dataReuploading}. This structure allows the network size to scale iteratively by adding qubits in a way that ensures the loss does not increase upon the introduction of a new qubit. This is achieved by initializing the training of the enlarged network with the optimized configuration obtained in the previous step with one fewer qubit. Such an approach can be regarded as a form of warm-start, a strategy that has recently been explored in the literature \cite{Warm_starts_Quantum,Warm_starts_PRX,warm_starts_arxiv}. While the architecture is general and can be extended to an arbitrary number of qubits, training pulse-level models from scratch is computationally demanding. Therefore, as a proof of concept, we begin with a single-qubit \gls*{qnn}, then extend it by adding a second qubit, ultimately training a two-qubit \gls*{qnn}. We envision this work also as an invitation to further develop and refine libraries that enable pulse-level programming.

We now outline the gate-level training scheme, which will subsequently be adapted to the pulse-level setting in the next section. We start from a single-qubit data re-uploading model with an $L$-layer structure of the form:
\begin{equation}\label{eq:QNN}
    U(\boldsymbol{\theta}^{(1)}_L) U(\boldsymbol{x}) \cdots U(\boldsymbol{\theta}^{(1)}_1) U(\boldsymbol{x}) |0\rangle,
\end{equation}
where each parameter vector is defined as
\begin{equation}
\boldsymbol{\theta}^{(1)}_l = (\theta^{(1)}_{l,1}, \theta^{(1)}_{l,2}, \theta^{(1)}_{l,3}) \in \mathbb{R}^3.
\end{equation}
Here, the superscript $(1)$ denotes the qubit index (in this case, the first qubit), and the subscript $l$ denotes the layer. The corresponding unitary is defined as
\begin{equation}\label{eq:one_qubit_param_gate}
U(\boldsymbol{\theta}^{(1)}_l) = R_Z(\theta^{(1)}_{l,1}) \, R_Y(\theta^{(1)}_{l,2}) \, R_Z(\theta^{(1)}_{l,3}),
\end{equation}
where $R_Y$ and $R_Z$ denote rotations around the $Y$ and $Z$ axes, respectively. This decomposition represents a generic $SU(2)$ unitary for a single qubit.

\begin{figure}[t!]
\centering
\includegraphics[width=\columnwidth]{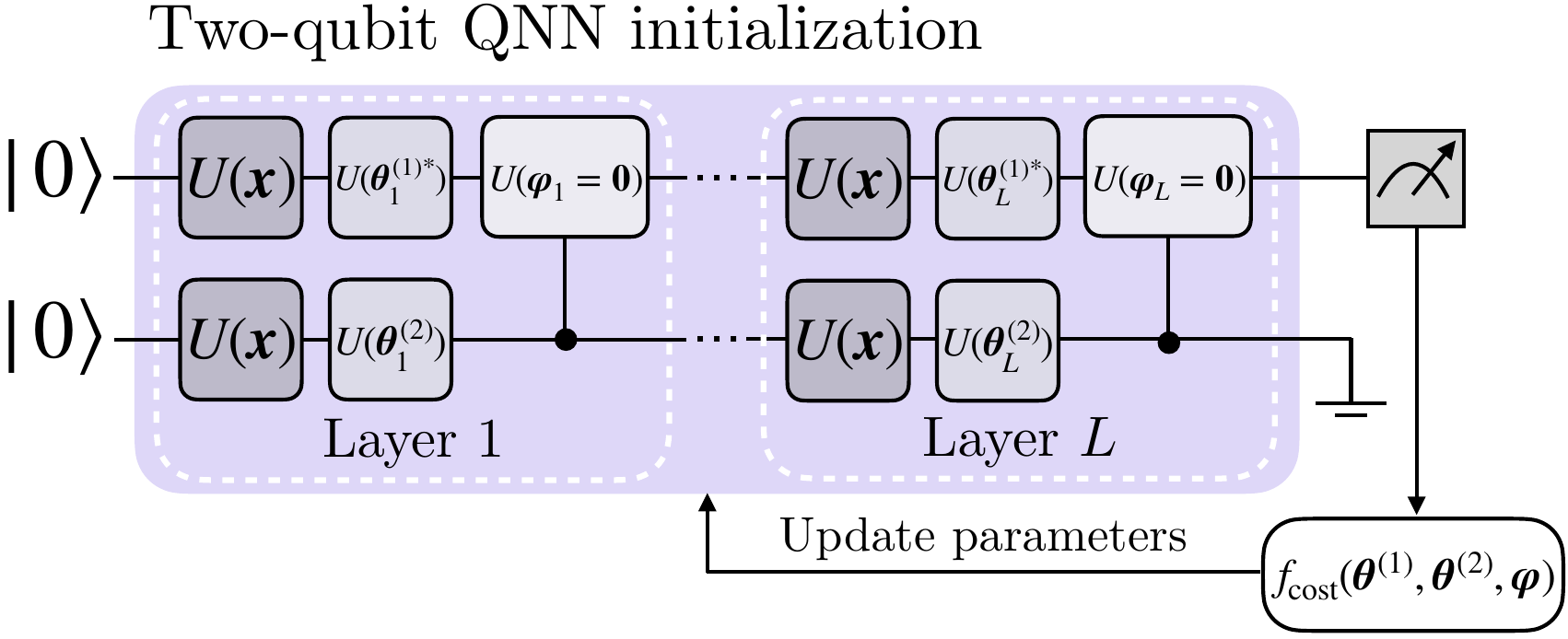}
  \caption{Initialization of the proposed two-qubit QNN training. The parameters of the first qubit are initialized with the optimized values obtained from the previous single-qubit QNN training. The parameters of the entangling gates are set to zero to ensure that both qubits start the training process in a product state. The parameters of the second qubit are initialized randomly. From this initialization, all parameters are subsequently updated to minimize the cost function.} 
\label{Fig:Pablo_multi_qubit_QNN}
\end{figure}

A fidelity-based cost function is employed, which drives inputs from class 1 towards the north pole of the Bloch sphere and inputs from class 0 towards the south pole, thereby implementing the binary classification task. Minimizing this cost yields the optimized parameters, denoted $\boldsymbol{\theta}^{(1)*} = \{\boldsymbol{\theta}^{(1)*}_1, \ldots, \boldsymbol{\theta}^{(1)*}_L\}$. Once the single-qubit \gls*{qnn} is trained, we extend the architecture by adding a second qubit. The new model mirrors the structure in Eq.~\eqref{eq:QNN}, now parameterized by $\boldsymbol{\theta}^{(2)}$, and includes controlled-$SU(2)$ unitaries acting on the first qubit with the second qubit as control. These two-qubit gates are parameterized by $\boldsymbol{\varphi} = \{\boldsymbol{\varphi}_l\}_{l=1}^L$.

The training procedure for the two-qubit \gls*{qnn} is as follows: the parameters of the single-qubit gates acting on the first qubit are initialized with the optimized values from the single-qubit \gls*{qnn}, $\boldsymbol{\theta}^{(1)} = \boldsymbol{\theta}^{(1)*}$; the parameters of the single-qubit gates on the second qubit are initialized randomly; and the two-qubit gate parameters are initialized at zero, effectively decoupling the two qubits at the start of training. Since the cost function is defined solely in terms of local measurements on the first qubit, this initialization guarantees that the cost function at the beginning of two-qubit training matches the final value achieved in the single-qubit case. During training, the two-qubit parameters deviate from zero only if this leads to further cost minimization, thereby allowing the second qubit to contribute adaptively. This is depicted in Fig.~\ref{Fig:Pablo_multi_qubit_QNN}.

\begin{figure}[t!]
    \centering
    \includegraphics[width=\columnwidth]{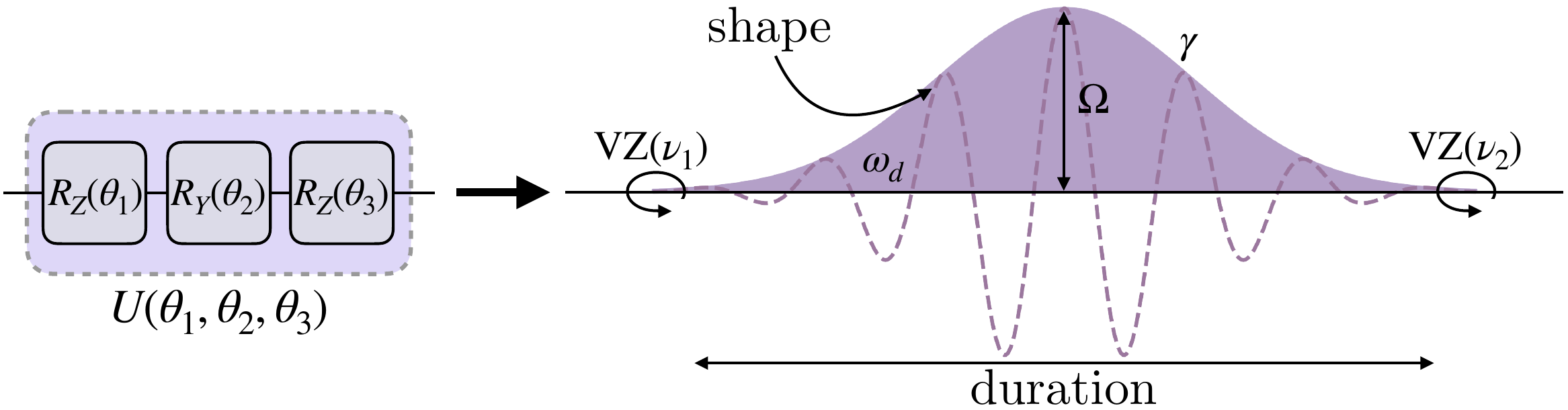}
    \hfill
    \caption{Pulse-based gate replacement. Each $Z$-rotation gate is implemented as a \gls*{vz} rotation, introducing neither physical errors nor additional duration. The fixed $Y$-rotation gate is replaced by a parameterized single-qubit pulse, enabling arbitrary rotations within the $XY$-plane. As a result, the trainable parameters consist of the pulse phase, pulse amplitude, and the two \gls*{vz} rotation angles. Note that the original $\nu_2$ parameter is linked to the pulse amplitude $\Omega$, the pulse shape $s(t)$, and the pulse duration.}
    \label{fig:gate_replaced_by_pulses}
\end{figure}

\section{Hardware-Adapted QML}\label{sec:proposed_model}

We now present a pulse-level formulation of the data re-uploading model, translating its standard \gls{vqc} architecture into a hardware-native implementation. Rather than constructing circuits from parameterized gates, we define native control pulses as the fundamental building blocks, allowing the model to interact directly with superconducting transmon hardware. This pulse-based formulation is designed to enhance expressivity, reduce calibration overhead, and improve robustness under realistic noise conditions, features particularly relevant in the NISQ regime. While our current demonstration focuses on data re-uploading, the framework provides a general methodology that can be extended to other variational QML architectures, enabling pulse-level implementations across diverse models.

Working at the pulse level affords enhanced expressivity without necessarily requiring deeper circuits~\cite{liang2024advantagesparameterizedquantumpulses}. Because pulses can be spectrally rich, featuring multiple frequency components, amplitude and phase modulation, or overlapping controls, they can implement transformations that would otherwise require many discrete gates. In effect, this compresses what would be a “deep” circuit into a shallower pulse schedule while retaining or even surpassing the representational power. Indeed, “infinitely deep” re-uploading limits emerge as the continuous-time limit of discrete blocks, giving pulse ansätze a potentially richer functional basis within the same coherence window~\cite{tao2024unleashingexpressivepowerpulsebased}.

This expressive freedom also brings significant advantages for noise resilience. By avoiding a fixed decomposition into discrete gate layers, pulse-level designs reduce total execution time and the number of concatenated, error-prone operations. They further bypass calibration errors inherent to pre-defined gate sets, mitigating decoherence and the accumulation of gate errors, and have been shown to improve empirical QML performance through pulse-efficient transpilation~\cite{Earnest_2021}.

Finally, one limitation that remains in VQC-based models, including data re-uploading, is the occurrence of barren plateaus, where cost-function gradients can vanish in large or deep circuits~\cite{barrenplateusinQML, barrenplateusinDR, McClean2018}. Pulse-level parameterization has been shown to partially alleviate this effect by producing smoother and more favorable optimization landscapes~\cite{barrenplateauspulses}, helping maintain trainability in realistic NISQ scenarios.

\begin{figure*}
    \centering
    \includegraphics[width=\linewidth]{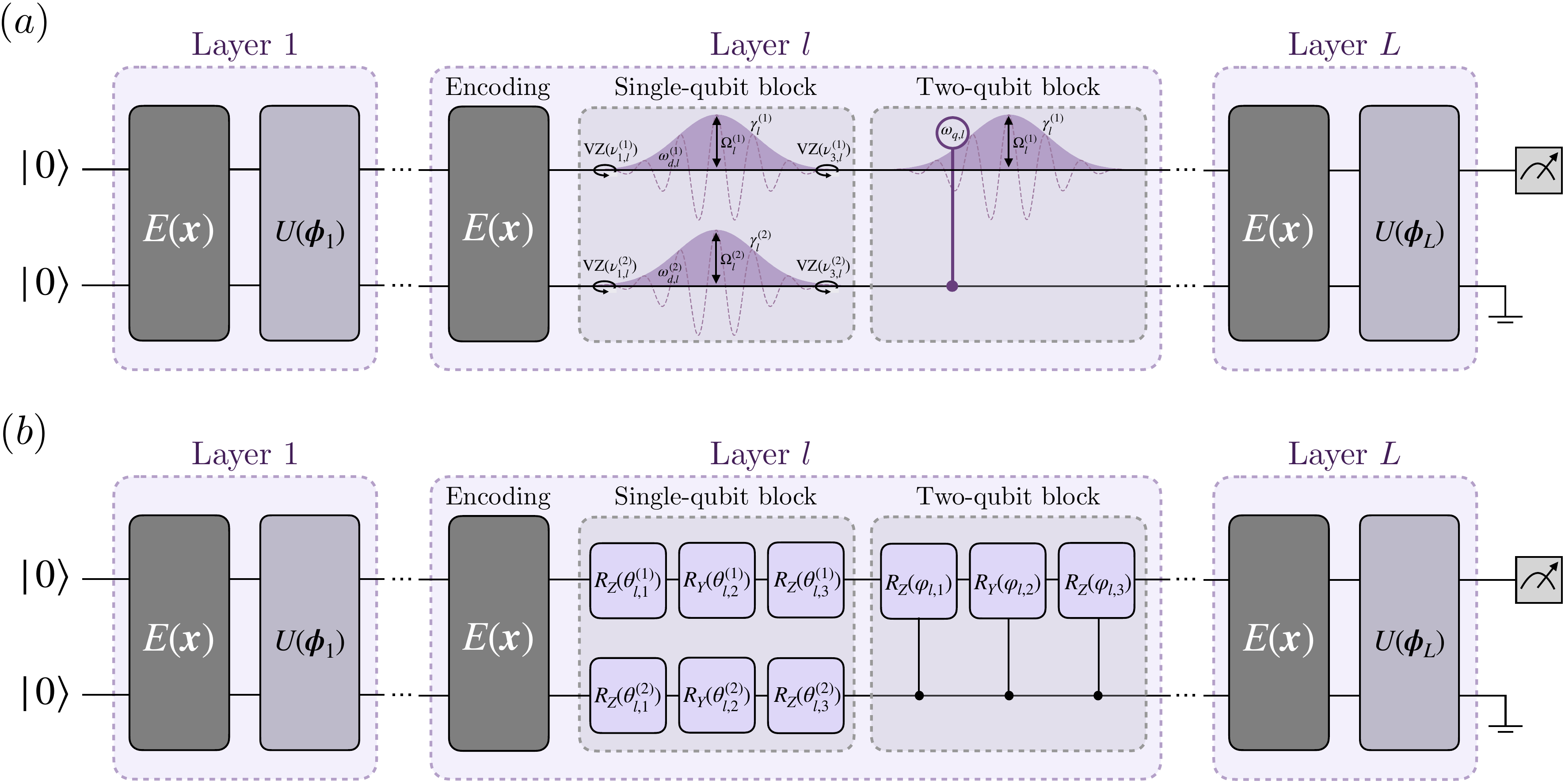}
    \caption{Schematic representation of the pulsed ansätze $(a)$ compared to the original gate-based ansätze $(b)$ of a data re-uploading model. As in the traditional approach, the pulsed \gls*{qnn} exhibits a layered structure, divided into three basic operations: an encoding block, a single-qubit parameterized block, and a multi-qubit variational gate scheme. In the pulse model, each of these parametric blocks corresponds to a sequence of parameterized pulses, with selected physical parameters left free for optimization. The encoding block remains general and can be replaced with any user- or problem-specific encoding. Here, $\boldsymbol{\phi_l}$ represents the full set of trainable parameters for layer $l$, corresponding to the pulse controls in $(a)$ and to the gate rotations in $(b)$.}
    \label{fig:pulsed_ansatz}
\end{figure*}

Taken together, these results suggest that pulse-level \gls*{qml} may not only be more expressive per unit time but can also be made intrinsically more noise-robust, provided that the pulse parametrization and control objectives are designed with explicit consideration of the device’s noise channels and calibration constraints. In this context, our approach extends variational quantum circuits to the pulse-programming level by replacing parameterized gates with native control pulses. Just as a \gls*{vqc} can be constructed from single- and multi-qubit blocks arranged in a layered or problem-specific structure, we define fundamental pulse operations whose physical parameters are left free for optimization. The key distinction is that, rather than tuning abstract circuit parameters with indirect physical meaning, we directly optimize the properties of the control pulses themselves—effectively communicating with the quantum device in its native language.

\subsection{Single-qubit parametrized quantum gates}
Analogous to single-qubit rotations being defined by three Euler angles, a general single-qubit operation can be physically realized via a control pulse, parameterized by attributes such as shape, amplitude, duration, phase, and frequency. While an SU(2) operation has only three degrees of freedom, these pulse parameters are often redundant—for instance, a pulse with amplitude $\Omega$ and duration $\tau$ may implement the same rotation as one with amplitude $\Omega/2$ and duration $2\tau$. However, this redundancy offers flexibility for optimizing other aspects, such as physical robustness, reduced leakage, or noise suppression~\cite{drag_pulses}. Although such optimizations are not the focus of this work, they highlight the advantages of adopting a pulse-level representation over rigid gate sequences.

Given this redundancy, training all these parameters does not improve expressivity but only increases computational cost. In our general pulse model, we therefore leave amplitude (which sets the rotation angle), phase (which primarily selects the rotation axis in the $XY$ plane), and frequency (which contributes to $Z$-rotations, qubit entanglement, and other effects~\cite{CRHamiltonian_errors}) as free control parameters, while keeping pulse duration and shape fixed. In total, we have three independent parameters for the three degrees of freedom that a SU(2) rotation has. The replacement of single-qubit gates with pulse-level operations is illustrated in Fig.~\ref{fig:gate_replaced_by_pulses}. 

For the single-qubit blocks in our ansatz, we fix the drive frequency by setting $\Delta_\omega = 0$ in Eq.~(\ref{eq:one_pulse_ham}). Resonant pulses are easier to implement and suppress unwanted off-resonant effects, and this choice does not restrict expressivity: arbitrary single-qubit rotations remain accessible by adjusting the pulse amplitude and phase together with the effectively free virtual-$Z$ rotations, which we keep explicitly as phase shifts.

Thus, each parameterized single-qubit pulse in our ansatz is defined by four parameters: two virtual-$Z$ rotation angles, and the amplitude and phase of the pulse:
\begin{equation}\label{eq:1qpulseblock}
    U_{\text{pulse}}^{(q)}(\nu_1, \nu_2, \Omega, \gamma)
    = V_Z^{(q)}(\nu_1)\, \mathcal{U}[(\Omega,\gamma)]\,
      V_Z^{(q)}(\nu_2),
\end{equation}
where $\mathcal{U}[(\Omega,\gamma)]$ is the time-evolution operator generated by the control Hamiltonian of Eq.~\ref{eq:one_pulse_ham} acting on qubit $q$, applied for a fixed duration $T$ and resonant with the qubit frequency $\omega_q$, with pulse phase $\gamma$ and amplitude $\Omega$, and $V_Z^{(q)}(\cdot)$ denotes a \gls*{vz} rotation on qubit $q$. The operator $\mathcal{U}[(\Omega,\gamma)]$  thus propagates the qubit state under the full driven Hamiltonian for the pulse duration, capturing both rotation and phase effects in the qubit’s Bloch sphere evolution.

\begin{table*}[t!]
\centering
\scriptsize
\begin{tabular}{@{}c c c c c c c c c c c c c c c@{}}
\toprule
\textbf{Qubit} & \textbf{T1 ($\mu$s)} & \textbf{T2 ($\mu$s)} & \textbf{Freq (GHz)} & \textbf{Anh. (GHz)} & \textbf{1Q Time (ns)} & \textbf{Coup. (GHz)} & \textbf{2Q Time (ns)} & \textbf{RO Err.} & \textbf{P(0|1)} & \textbf{P(1|0)} & \textbf{Rz Err.} & \textbf{SX Err.} & \textbf{X Err.} & \textbf{ECR Err.} \\ 
\midrule
1 & 180 & 180 & 4.8 & -0.31 & 300 & \multirow{2}{*}{0.013} & \multirow{2}{*}{660} & 0.0337 & 0.0215 & 0.0459 & 0 & 0.000187 & 0.000187 & 0.00431 \\
2 & 310 & 250 & 4.6 & -0.31 & 300 &  &  & 0.0256 & 0.0176 & 0.0337 & 0 & 0.000367 & 0.000367 & 0.00431 \\ 
\bottomrule
\end{tabular}
\caption{Statistics for Brisbane qubits 1 and 2, including readout and gate errors. The data are extracted from the official publicly available device datasheet~\cite{brisbane_device}.}
\label{tab:brisbane_tab}
\end{table*}

It is important to note that this single-qubit gate replacement does not enhance expressivity, as it remains an arbitrary SU(2) transformation as before. Nonetheless, it may facilitate faster or more effective learning in the model.

\subsection{Two-qubit parametrized quantum gates}
Designing two-qubit interactions at the pulse level poses a particular challenge. While general controlled SU(2) operations are often assumed in gate-based models, realizing such operations with pulses typically requires multiple calibrated signals, which increases the total execution time and exposes the system to additional sources of noise and decoherence~\cite{nonlinearerrorNISQ}. For instance, a basic CNOT gate may require two or more control pulses, depending on the hardware-specific transpilation~\cite{quantumGuide}, as displayed in Fig.~\ref{fig:transmon_gates}{\color{blue}$(b)$}.

To ensure physical feasibility and minimize experimental overhead, we adopt a more hardware-aligned strategy inspired by the \gls*{cr} gate. Specifically, we entangle two qubits using a single microwave pulse applied to one qubit at or near the resonance frequency of the other. The pulse’s amplitude and phase are treated as trainable parameters, and an additional frequency detuning term is introduced to allow the model to explore a continuum of entangling interactions beyond the standard $XZ$-rotation. This added flexibility enables the circuit to dynamically modulate the strength and nature of entanglement, and to suppress it altogether if not required by the learning task.
As a result, each two-qubit interaction is characterized by three learnable parameters of the pulse: the amplitude, the phase, and the frequency detuning. 

\subsection{Encoding blocks}
While the encoding block could also be implemented using parameterized pulses, we leave this extension as future work. In this study, our primary focus is on investigating how introducing pulse-level operations affects the expressivity and noise resilience of the quantum model.

Exploring pulse-based encoding could further enhance the flexibility of the model by allowing the data representation itself to adapt to the native control landscape of the hardware. However, analyzing this aspect is beyond the scope of the current work, where we aim to isolate and characterize the impact of pulse-level optimization on the model’s performance and robustness in the presence of realistic noise.

\begin{figure*}[t]
    \centering
    \includegraphics[width=\linewidth]{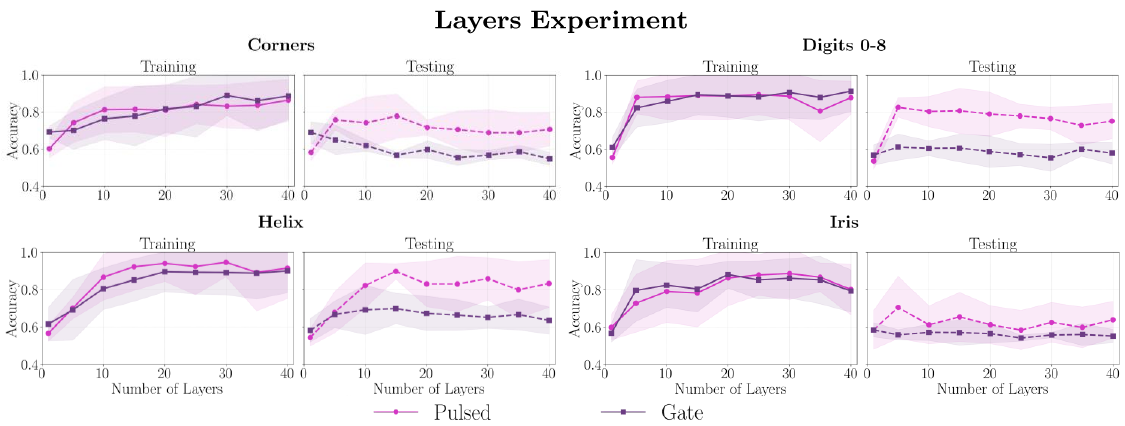}
    \caption{Performance comparison of the traditional gate-based QNN (Gate, purple) and the proposed pulse-based QNN (Pulsed, pink) across a varying number of layers using a fixed two-qubit architecture. The pulse model achieves comparable or superior test accuracy with fewer layers and exhibits a delayed onset of overfitting.}
    \label{fig:results_layers}
\end{figure*}

\section{Results}\label{sec:results}
We now assess the performance of the proposed pulse-based protocol through numerical simulations, focusing on its ability to maintain generalization capabilities while demonstrating robustness to noise across tasks.

\subsection{Experimental setup}
As a proof of concept for our hardware-adapted design, we numerically implement the data re-uploading model, explained in Section~\ref{subsec:qml}, and use it for binary classification tasks. Although our experiments focus on this particular architecture, the proposed approach is general and can be extended to multi-class settings or kernel-based learning by appropriately modifying the cost function and readout strategy.

In the data re-uploading model introduced in Section~\ref{sec:proposed_model}, the encoding stage is implemented using standard gate-based operations, while the variational component is replaced by a pulse-based version. This last part consists of single- and multi-qubit layers, where parameterized gates are substituted with native control pulses. Single-qubit SU(2) rotations are realized through pulse-parameterized operations, and entangling layers through single-pulse, cross-resonance–inspired interactions with tunable amplitude, phase, and detuning. The control fields used in our pulsed model follow a sinusoidal form of constant amplitude, $\Omega \sin(\omega_d t + \gamma)$. With this pulse shape, the effective rotation angle of a single-qubit operation scales linearly with the product $\Omega T$, where $T$ denotes the pulse duration. The circuit layout is shown in Fig.~\ref{fig:pulsed_ansatz}.

Additionally, to improve learning flexibility, we relax the constraint that class labels correspond strictly to computational basis states. Instead, the target states are defined as general single-qubit states 
\begin{align}
    \ket{s_0(\theta, \varphi)} &= \cos(\theta) \ket{0} + e^{i\varphi} \sin(\theta )\ket{1}\\
    \ket{s_1(\theta, \varphi)} &=  -\sin(\theta) \ket{0} + e^{i\varphi} \cos(\theta) \ket{1}
\end{align}
with both $\theta$ and $\varphi$ treated as learnable parameters. This formulation allows the model to dynamically adjust the decision boundary during training, potentially accelerating convergence and enhancing classification performance.

As the training cost function, we employ the fidelity-based loss introduced in~\cite{dataReuploading}, defined as
\begin{equation}
f(\bm X,\bm Y) = \sum_{(x,y)\in \bm X \times \bm Y} \frac{\left(1 - \langle \psi(x, \bm \theta) | \psi_l(y) \rangle \right)^2}{|\bm X|},
\end{equation}
where $\bm X$ denotes the input samples from the training set and $|\bm X|$ the number of training points, $\bm Y$ their corresponding labels, $\ket{\psi(x, \bm \theta)}$ is the quantum state produced by the \gls*{qnn} for input $x$ and parameters $\bm \theta$, and $\ket{\psi_l(y)}$ is the target state associated with label $y$. 

We evaluate our models on a diverse set of binary classification tasks comprising both real-world and synthetic datasets. Specifically, we consider the MNIST handwritten digits dataset~\cite{mnist_digits}, where the classification task consists of distinguishing digits 0 and 8; and the Iris dataset~\cite{iris_53}, reduced to the two most similar classes, \textit{Iris versicolor} and \textit{Iris virginica}. Additionally, we introduce two synthetic datasets designed to probe different types of geometric decision boundaries: \textit{Corners}, in which the vertices of a cube belong to one class while the points within the cube belong to the other class (Fig.~\ref{fig:datasets}, Left), and \textit{Helix}, consisting of two intertwined helices with a relative phase shift of $\pi/2$ (Fig.~\ref{fig:datasets}, Right).

To ensure compatibility with the three-dimensional quantum embedding, MNIST images are first projected onto a three-dimensional feature space using classical PCA. For the MNIST digits, Corners, and Helix datasets, we randomly select 300 samples for training and 100 samples for testing. Due to the limited size of the Iris dataset, which contains only 100 samples in the selected binary setting, we instead employ a split of 70 training samples and 30 testing samples. All experiments are repeated over five different random seeds.

To assess the performance of the proposed model under realistic conditions, we simulate its execution on IBM’s superconducting quantum processor \texttt{ibm\_brisbane}, using publicly available device parameters and control constraints~\cite{brisbane_device}. 

Due to the high computational cost of simulating microwave-driven qubit dynamics with noise~\cite{footnote_codeimplementation}, we restrict our analysis to two-qubit configurations. Although extending this study to larger multi-qubit systems represents a natural and important direction for future work, the considerable computational demands of pulse-level noisy simulations currently limit the feasible scope of our analysis. The simulations presented in this work were performed using a custom extension of the \texttt{pennylane}~\cite{bergholm2022pennylaneautomaticdifferentiationhybrid} framework developed specifically for this research. Nevertheless, the use of more mature and professionally maintained simulation frameworks, potentially incorporating optimized numerical methods and high-performance computing support, may facilitate the extension of this analysis to larger quantum systems in future studies.

\begin{figure*}[t!]
    \centering
    \includegraphics[width=\linewidth]{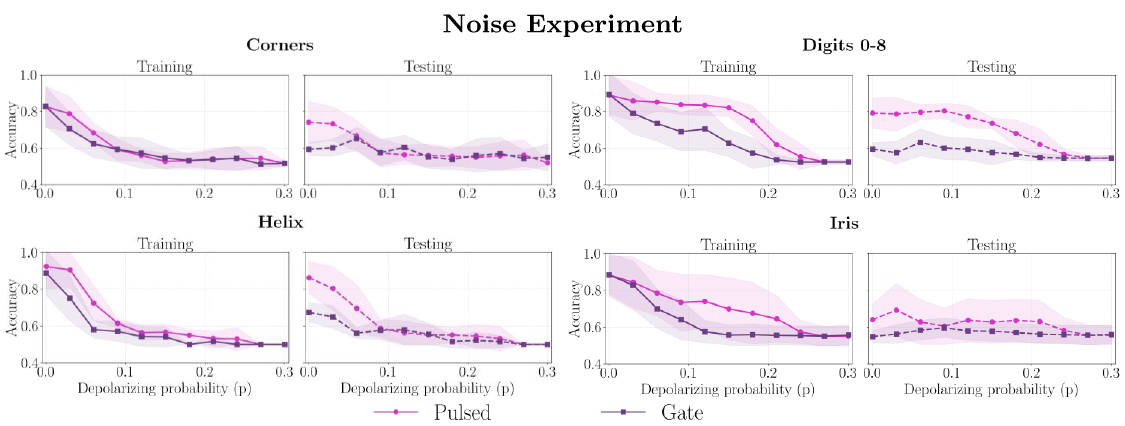}
    \caption{Performance comparison of the traditional gate-based QNN (Gate, purple) and the proposed pulse-based QNN (Pulsed, pink) under varying depolarizing channel noise probabilities, using a fixed two-qubit, 20-layer \gls*{qnn} architecture. The pulse model retains useful classification accuracy across a wider noise range before converging to the random-guessing baseline.}
    \label{fig:results_noise}
\end{figure*}

\subsection{Generalization performance evaluation}
The central finding of this experiment is that pulse-level optimization delays the onset of overfitting, enabling higher test accuracy at equivalent or shallower depths. We compare the generalization performance of the proposed pulsed model with its gate-based counterpart by analyzing both training and test accuracy as a function of the number of re-uploading layers, as shown in Fig.~\ref{fig:results_layers}. While increasing the number of layers typically enhances model expressivity, leading to improved training performance, it also increases the risk of overfitting the training data, which can negatively impact the ability to generalize to unseen data.\\ \\

Both models generally benefit from increased circuit depth, with 
training accuracy improving monotonically as the number of layers 
grows. However, despite reaching comparable training performance 
at large depths, the two parameterizations display markedly 
different generalization behaviors and exhibit markedly different overfitting  behavior. The advantage of the  Pulsed model becomes particularly evident when examining the test 
accuracy curves. In this sense, the gate-based model develops a pronounced train–test gap for $L > 10$, especially on the \textit{Helix} and \textit{Digits} datasets, where test accuracy stagnates or deteriorates despite near-perfect training performance—indicating that additional depth primarily enables memorization rather than feature extraction. In contrast, the pulse-based model maintains a closer correspondence between training and testing accuracy across a wider range of depths: the onset of overfitting is consistently delayed, and competitive test accuracy is frequently achieved with fewer layers than the gate-based model requires. This suggests a more efficient use of model complexity, a particularly attractive property for near-term devices where circuit depth is a critical resource.

Overall, these results indicate that pulse-level optimization can provide a more favorable trade-off between expressivity and generalization compared with conventional gate-based parameterizations. By directly optimizing hardware-native control pulses, the model appears capable of learning effective representations with shallower architectures while remaining less susceptible to overfitting as the depth increases.

\subsection{Noise resilience evaluation}
The key observation is that the pulse-based model retains classification performance well into the moderate-noise regime ($p \lesssim 0.15$), whereas the gate-based model degrades substantially earlier. We fix the number of layers at 20, keep all noise parameters from Table~\ref{tab:brisbane_tab} constant, and vary only the depolarizing channel probability $p$ from $0$ to $0.3$ (Fig.~\ref{fig:results_noise}).

As expected, both models exhibit a monotonic degradation in 
accuracy as $p$ increases, with performance converging toward the 
random-guessing baseline ($\sim 0.5$) in the high-noise regime. 
However, the two architectures differ notably in their resilience 
to depolarizing noise, with the Pulsed model demonstrating a 
consistent advantage throughout the low-to-moderate noise range.

This advantage is most pronounced on the \textit{Digits} and 
\textit{Iris} datasets, where the gap between the two 
approaches is substantial and persists across both training and 
testing curves. The fact that the improvement appears on both 
curves is particularly significant: it rules out the possibility 
that the Pulsed model's advantage is merely a byproduct of 
reduced overfitting, and instead points to a genuine structural 
robustness of the pulse-based ansatz to depolarizing noise. On 
the \textit{Corners} and \textit{Helix} datasets the gap is 
comparatively smaller, yet the Pulsed model still maintains 
higher or comparable accuracy throughout most of the noise range. 
Notably, on \textit{Digits}, the improvement 
in noise resilience is accompanied by a systematic increase in 
test accuracy relative to the gate-based model, further 
confirming the generalization advantage identified in the 
previous experiment.

For $p \gtrsim 0.2$, the performance of both models saturates 
near the random-guessing baseline, indicating that beyond this 
threshold the quantum information is too severely disrupted for 
either architecture to extract meaningful signal. Additionally, 
the Pulsed model exhibits wider confidence bands in several 
benchmarks, which may reflect a higher sensitivity of its 
continuous parameterization to stochastic noise realizations, 
an aspect that suggests further investigation.

In summary, the pulse-based model exhibits a structurally more robust response to depolarizing noise across all benchmarks. The fact that its advantage persists on both training and test curves indicates that the improvement stems from a genuine representational benefit of the pulse ansatz, rather than from regularization effects alone. We note that the depolarizing probabilities explored here exceed typical single-gate error rates; however, they provide a controlled stress test that reveals relative robustness differences that would otherwise require prohibitively large multi-qubit simulations. Extending these observations to larger system sizes remains an important direction for future work.

\section{Conclusions}\label{sec:conclusions}
We have introduced a general methodology for translating gate-based variational QML architectures into pulse-level implementations, and demonstrated it on a data re-uploading QNN simulated under realistic transmon noise.

Our experiments reveal three consistent findings. First, the pulse-based model achieves competitive or superior test accuracy with fewer re-uploading layers, and its train–test gap grows more slowly with depth than in the gate-based counterpart, indicating improved generalization. Second, under increasing depolarizing noise the pulse model sustains useful accuracy across a wider range of error probabilities, with the advantage appearing on both training and test curves—pointing to a structural, not merely regularization-based, benefit in noise resilience. Third, these trends hold on four qualitatively different binary classification benchmarks, providing evidence that the observations are not artifacts of a single dataset.\\

These findings suggest that operating at the native pulse level can yield a more favorable expressivity–robustness trade-off than conventional gate parameterizations, supporting pulse-native QML as a promising direction for NISQ-era devices. From a practical standpoint, the reduced effective circuit depth and tighter coupling to hardware dynamics may translate into lower calibration overhead and better utilization of limited coherence times on real processors.

We acknowledge two main limitations. First, computational cost restricted our simulations to two-qubit systems; scaling to larger registers will require community investment in pulse-level simulation infrastructure and differentiation toolkits. Second, the encoding stage remains gate-based in the present work; extending it to the pulse level is a natural next step that could further enhance hardware alignment, though it introduces additional experimental confounders that merit a dedicated study.

A particularly compelling direction for future research is a formal analysis of the loss landscape of pulse-parameterized models. The wide dispersion in final accuracy observed across seeds (Figures~\ref{fig:results_layers} and~\ref{fig:results_noise}) strongly suggests a rugged optimization landscape, raising the question of whether specialized optimizers—such as landscape-aware or noise-adaptive strategies—could unlock the full capacity of these architectures. Equally important is expanding the evaluation beyond two-qubit systems; while this scaling is currently constrained by simulation cost, community-driven efforts toward pulse-level simulation infrastructure and specialized training frameworks~\cite{franz2026softwarequantummachinelearning} could significantly alleviate these limitations and enable validation at practically relevant scales.

\section*{Code \& Data Availability}
The code and data for all experiments is available at \href{https://github.com/nacedob/Pulsed-Data-Reuploading-Quantum-Models}{https://github.com/nacedob/Pulsed-Data-Reuploading-Quantum-Models}.

\begin{acknowledgments}
We thank Adrián Pérez-Salinas for helpful comments on early stages of the project. PRG, PGA and JGC acknowledge support from HORIZON-CL4-2022-QUANTUM01-SGA project 101113946 OpenSuperQ-Plus100 of
the EU Flagship on Quantum Technologies, the Spanish Ram\'on y Cajal Grant RYC-2020-030503-I, and the “Generaci\'on de Conocimiento” project Grant No. PID2021-125823NA-I00 funded by MICIU/AEI/10.13039/501100011033, by “ERDF Invest in your Future” and by FEDER EU. PGA acknowledges support from UPV/EHU Ph.D. Grant No. PIFG 22/25. We also acknowledge support from the Basque Government through Grants No. IT1470-22, the Elkartek project KUBIBIT - kuantikaren berrikuntzarako ibilbide teknologikoak (ELKARTEK25/79), and from the IKUR Strategy under the collaboration agreement between Ikerbasque Foundation and BCAM on behalf of the Department of Education of the Basque Government. This work has also been partially supported by the Ministry for Digital Transformation and the Civil Service of the Spanish Government through the QUANTUM ENIA project call – Quantum Spain project, and by the European Union through the Recovery, Transformation and Resilience Plan – NextGenerationEU within the framework of the Digital Spain 2026 Agenda.

\end{acknowledgments}


\appendix
\section{Quantum Optimal Control\label{sec:Appendix QOC}}
\subsection{The Transmon Qubit Hamiltonian}

To understand transmon's physics, we start with the LC circuit, equivalent to a quantum harmonic oscillator with Hamiltonian~\cite{quantumGuide}
\begin{equation}
    \hat H_{LC} = \frac{\hat Q^2}{2C} + \frac{\hat \Phi^2}{2L},
\end{equation}
where $\hat Q$ and $\hat \Phi$ are conjugate operators. The spectrum is harmonic, preventing isolation of a two level system. To introduce anharmonicity, the linear inductor is replaced with a Josephson junction, forming the Cooper Pair Box (CPB) qubit. This qubit can be described by the Hamiltonian
\begin{equation}
    \hat H = 4E_C (\hat n - n_g)^2 - E_J \cos \hat \varphi,
\end{equation}
with $E_C$ the charging energy, $E_J$ the Josephson energy, $\hat n$ the Cooper-pair number operator, and $\hat \varphi$ the superconducting phase, and $n_g$ is the offset charge bias.

In practice, transmons operate in the regime $E_J \gg E_C$, where the phase $\varphi$ is localized and the system is stable under charge fluctuations. Expanding the cosine to fourth order,
\begin{equation}
    \cos \varphi \simeq 1 - \frac{\varphi^2}{2} + \frac{\varphi^4}{24},
\end{equation}
the quadratic term gives the harmonic oscillator contribution, while the quartic term provides the required anharmonicity.

Expressing $\hat n$ and $\hat \varphi$ in terms of ladder operators and neglecting constants and fast-oscillating terms, one obtains the approximate Hamiltonian
\begin{equation}
    \hat H \simeq \tilde{\omega} \hat a^\dagger \hat a + \frac{\delta}{2}(\hat a^\dagger \hat a)^2,
\end{equation}
with $\tilde{\omega} = \sqrt{8E_C E_J}$ and $\delta \simeq -E_C$. This yields non-equidistant levels:
\begin{equation}
    \hat H_{\text{transmon}} = \sum_j \left[\left(\omega_0 - \frac{\delta}{2}j\right) + \frac{\delta}{2} j^2 \right]\ket{j}\bra{j}.
\end{equation}

By truncating to the two lowest states $\{\ket{0}, \ket{1}\}$, one arrives at the effective qubit Hamiltonian
\begin{equation}
    \hat H_{\text{transmon},2} \simeq -\frac{\omega_0}{2}\,\sigma_z,
\end{equation}
where higher levels remain weakly accessible due to finite anharmonicity. For improved modeling, one can include the anharmonic correction term $\tfrac{\alpha}{2}(\hat a^\dagger \hat a - \mathbb{I})\hat a^\dagger \hat a$ ~\cite{quantumGuide}.

\subsection{Driven transmon: effective qubit Hamiltonian}

The control of the effective two-level system just described, can be addressed by adding an external microwave drive field, capacitively coupled to the circuit. The qubit-field interaction can be modelled by the Hamiltonian~\cite{qiskit_nb_transmon}
\begin{equation}
\begin{split}
\hat H \;=& \; -\frac{\omega_0}{2}\hat\sigma_z \;-\; \Omega\big(\,e^{i\omega_d t}\hat\sigma^+ + e^{-i\omega_d t}\hat\sigma^-\,\big)
\;\\
=&\; -\frac{\omega_0}{2}\hat\sigma_z \;-\; \Omega\big(\cos(\omega_d t)\hat\sigma_x - \sin(\omega_d t)\hat\sigma_y\big).
\end{split}
\end{equation}
where counter rotating terms have been dropped by means of the RWA approximation assuming Rabi frequencies much smaller than anharmonicity and qubit frequencies. If we further consider introducing a phase in the drive frequency, we can tune rotation in the $XY$ plane
\begin{equation}
\begin{split}
    H_t = -\frac{\omega_0}{2}\sigmaz  + \Omega(t) \big[&\cos(\omega_d t - \gamma) \sigmax \\
    &+ \sin(\omega_d t - \gamma) \sigmay\big],
\end{split}
\end{equation}
where $\gamma$ sets the phase of the drive. In the rotating frame (and under the RWA), the Hamiltonian reduces to
\begin{equation}
    H \simeq  \frac{\Omega(t)}{2} \big(\cos\gamma \,\sigmax + \sin\gamma\, \sigmay \big),
\end{equation}
so the drive axis is controlled by $\gamma$, with the rotation angle $\theta = \int \Omega(t) \, dt$.

\subsection{Virtual $Z$ gates}\label{ap:sec:vzgates}

While direct $Z$ rotations can be performed via detuned pulses, a more efficient approach exploits \emph{Virtual $Z$} (\gls{vz}) gates~\cite{vitualZgates}. These gates are implemented purely in software: they adjust the reference frame of the qubit rather than applying a physical pulse, making them effectively instantaneous and error-free.

A rotation about an arbitrary axis in the $XY$ plane, $RX_\theta^{(\gamma)}$, can be decomposed as
\begin{equation}
    RX_\theta^{(\gamma)} = RZ_{-\gamma} \, RX_\theta \, RZ_\gamma,
\end{equation}
showing that a phase-shifted pulse is equivalent to a physical $X$ rotation sandwiched between $Z$ rotations. Consequently, any $Z$ rotation can be applied virtually by adjusting the phase of subsequent pulses, avoiding a real $Z$ pulse, as depicted in Fig.~\ref{fig:transmon_gates}{\color{blue}$(a)$}.  

This property is particularly useful since any single-qubit gate can be decomposed into a combination of $RX$ and $RZ$ rotations (Euler angles), e.g.
\begin{equation}
\begin{aligned}
    U(\theta, \phi,& \lambda) =\\ &
    RZ_\phi \, RX_{\pi/2} \, RZ_{\pi-\theta} \, RX_{\pi/2} \, RZ_{\lambda-\pi/2}.
\end{aligned}
\end{equation}
Therefore, using virtual $Z$ gates allows implementing arbitrary single-qubit operations with minimal pulse overhead and reduced error accumulation, as the final $Z$ rotation can often be absorbed into the frame of subsequent gates or the measurement basis.

\subsection{Connecting transmon qubits}

Qubits must be connected to enable two-qubit operations. While many coupling schemes exist, a highly exploited architecture are the fixed frequency couplers, used in IBM devices~\cite{quantumGuide}, which avoid the need for tunable qubit frequencies  or coupling elements. In general, capacitive coupling introduces an interaction term of the form $H_{\rm int} \propto \hat n_1 \hat n_2$, which in terms of ladder operators leads to a Jaynes–Cummings-type Hamiltonian between the qubits and the coupler or resonator.

In the dispersive regime, where the qubit–resonator couplings $g_{qr}$ are small compared to the detunings $\Delta_{q,r} = \omega_q - \omega_r$, the resonator degrees of freedom can be adiabatically eliminated~\cite{couplingTermAdiabatic}. Projecting onto the zero-resonator-excitation subspace yields an effective two-qubit Hamiltonian:
\begin{equation}\label{eq_ap:2qubitham}
    \hat H_{\rm eff} \simeq -\sum_{q=1,2} \frac{\omega_q}{2} \hat\sigma_q^z + J (\hat\sigma_1^x \hat\sigma_2^x + \hat\sigma_1^y \hat\sigma_2^y),
\end{equation}
with the effective coupling strength
\begin{equation}
    J = \frac{g_1 g_2 (\omega_1 + \omega_2 - 2 \omega_r)}{2 (\omega_1 - \omega_r)(\omega_2 - \omega_r)}.
\end{equation}

This expression is independent of the specific details of the coupler, highlighting that different architectures converge to the same effective qubit–qubit interaction. Such a representation is sufficient for analyzing two-qubit gates in software, for example when counting the number of CNOTs required to implement a SWAP.

A common two-qubit gate implemented in this architecture is the cross-resonance (\gls*{cr}) gate. The \gls*{cr} gate is a fundamental two-qubit operation in fixed-frequency transmon architectures. While its use for entangling qubits is described in the main text, several practical aspects merit discussion in an appendix. Driving one qubit at the frequency of another produces an effective $XZ$-type interaction. The resulting Hamiltonian includes higher-order contributions that can generate small off-diagonal rotations and residual $ZZ$ couplings, typically mitigated through echo sequences or careful calibration.  

The amplitude, duration, and phase of the control pulse must be tuned to maximize fidelity and minimize leakage to higher energy levels. In particular, AC-Stark shifts induced on the control qubit are compensated using virtual Z corrections, preserving the intended computational phases. Smooth pulse shaping, such as Gaussian or DRAG-inspired envelopes, further reduces spectral leakage and mitigates unwanted excitations outside the computational subspace.  

From a modeling perspective, it is often sufficient to consider the leading-order effective Hamiltonian
\begin{equation}
    \hat H_{\rm CR} \sim \mu\ \sigmaz \otimes \sigmax,
\end{equation}
which captures the dominant entangling behavior. Higher-order corrections can be treated as calibration subtleties. These practical considerations highlight why the CR gate is widely adopted: it provides robust two-qubit entanglement while allowing efficient software-level phase corrections and pulse optimizations.

\section{Noise Modelling}\label{sec:noise}

This section details the techniques used to model the noise of a real superconducting device, specifically the IBM Brisbane architecture. This modelling is essential for evaluating the performance of our proposed model under realistic conditions. The noise model is constructed using the standard framework of quantum channelss~\cite{Georgopoulos21, Piskor2025}, which we summarize below, followed by its parameterization based on the hardware characteristics of the target device.

\subsection{Quantum Channels and Kraus Operators}
Noise in quantum systems is typically described using the formalism of \emph{quantum channels}, which are completely positive trace-preserving (CPTP) maps acting on density matrices. A convenient representation of these channels is the \emph{operator-sum representation}, also known as the Kraus representation. In this formalism, the action of a noisy channel $\mathcal{E}$ on a density matrix $\rho$ is given by \cite{Nielsen_Chuang_2010}
\begin{equation}
\mathcal{E}(\rho) = \sum_{k} K_k \rho K_k^\dagger,
\end{equation}
where ${K_k}$ are the Kraus operators associated with the channel, satisfying the completeness relation $\sum_k K_k^\dagger K_k = I$. This framework provides a powerful tool for modelling the various noise processes that affect superconducting qubits due to interactions with their environment.

\subsection{Superconducting Noise Contributions}

\subsubsection{Depolarizing Channel}
The depolarizing channel represents a stochastic process where the qubit state is replaced by the maximally mixed state with probability $p$. For a single qubit, the Kraus operators are:
\begin{equation}\label{eq:depolarizing_channel_kraus}
\begin{aligned}
K_0 &= \sqrt{1 - p} I, \quad &K_1 = \sqrt{\frac{p}{3}} \hat\sigma^X, \\
K_2 &= \sqrt{\frac{p}{3}} \hat\sigma^Y, \quad &K_3 = \sqrt{\frac{p}{3}} \hat\sigma^Z.
\end{aligned}
\end{equation}
Here, $p \in [0,1]$ denotes the probability of a random Pauli error. In the context of superconducting qubits, depolarization is often used as a simplified model that captures the combined effect of various uncharacterized error sources.

\subsubsection{Amplitude Damping}
The amplitude damping channel models energy relaxation, where a qubit decays from the excited state $\ket{1}$ to the ground state $\ket{0}$. The Kraus operators are:
\begin{equation}\label{eq:amplitude_damping_kraus}
K_0 =
\begin{pmatrix}
1 & 0 \\
0 & \sqrt{1-\gamma}
\end{pmatrix}, \quad
K_1 =
\begin{pmatrix}
0 & \sqrt{\gamma} \\
0 & 0
\end{pmatrix},
\end{equation}
with $\gamma \in [0,1]$ denoting the decay probability. This process is directly linked to the finite energy lifetime of excitations in superconducting circuits.

\subsubsection{Phase Damping}
The phase damping channel models the loss of quantum coherence (dephasing) without energy loss. The Kraus operators are:
\begin{equation}\label{eq:phase_damping_kraus}
K_0 =
\begin{pmatrix}
1 & 0 \\
0 & \sqrt{1-\lambda}
\end{pmatrix}, \quad
K_1 =
\begin{pmatrix}
0 & 0 \\
0 & \sqrt{\lambda}
\end{pmatrix},
\end{equation}
where $\lambda \in [0,1]$ is the dephasing probability. This noise originates from low-frequency fluctuations, such as flux or charge noise.

\subsubsection{State Preparation and Measurement (SPAM) Errors}
\label{subsub:spam_errors}
In addition to dynamical noise during circuit execution, real quantum devices suffer from errors in state preparation and measurement (SPAM). These are critical to model as they directly affect the initial condition and the final readout of any computation. SPAM errors are typically characterized by:
\begin{itemize}
\item \textbf{Preparation Error}: The initial state $\ket{0}$ may be prepared imperfectly. This is often modelled as a classical probability $p_{\text{prep}}$ that the qubit is initialized in the $\ket{1}$ state instead of $\ket{0}$.
\item \textbf{Measurement Error}: The process of measuring a qubit in the computational basis can misclassify the outcome. This is described by a confusion matrix. For a single qubit, the probability of correctly measuring $\ket{0}$ is $p(0|0)$, and the probability of incorrectly measuring $\ket{1}$ when the state was $\ket{0}$ is $p(1|0) = 1 - p(0|0)$. Similarly, the probability of correctly measuring $\ket{1}$ is $p(1|1)$.
\end{itemize}

\subsection{Modelling the IBM Brisbane Device}
\label{sub:modelling_brisbane}
To validate our model with realistic numerical simulations, we parameterize the noise channels described above using the specifications of the IBM Brisbane device. As the simulation of larger systems is computationally expensive, we restrict our study to two-qubit models, which are sufficiently complex to demonstrate the potential of our contributions.

The relevant parameters for the selected qubits, such as decoherence times ($T_1$, $T_2$) and gate error rates, are obtained from the device's backend properties, summarized in Table~\ref{tab:brisbane_tab}. For a gate of duration $t$, the probabilities $\gamma$ and $\lambda$ associated with amplitude and phase damping channel parameters are computed as~\cite{Nielsen_Chuang_2010}:
\begin{equation}\label{eq:amplitude_damping_decoherence_time}
\gamma = 1 - e^{-t/T_1}, \quad \lambda = 1 - e^{-t/T_2}.
\end{equation}
These values are subsequently used to construct the corresponding Kraus operators from Eqs.~\eqref{eq:amplitude_damping_kraus} and \eqref{eq:phase_damping_kraus}.

Depolarizing error probabilities $p$ are not directly reported in the device datasheet. To estimate them, we average the measured error rates of the native single-qubit gates, specifically the $X$ and $\sqrt{X}$ gates. For two-qubit operations, the depolarizing probability is taken directly from the characterized error of the ECR gate.

SPAM errors are incorporated using the measured probabilities from device calibration. In our simulations, these errors are modeled by applying a classical probabilistic flip to the ideal initial state and final measurement outcomes, according to the characterized error rates.

This comprehensive model, combining incoherent decoherence, coherent gate errors, and SPAM errors, allows us to simulate the execution of our pulse-based algorithms on a realistic noisy quantum processor.

Once the basic Kraus operators for the noise channels are determined, they are applied at the appropriate points following~\cite{Georgopoulos21}. Specifically, depolarizing, amplitude damping, and phase damping channels are applied after each electromagnetic control pulse or single-qubit gate. For two-qubit operations, the generic controlled SU(2) rotation is first decomposed into the native device gates, and the corresponding single-qubit errors, as well as the ECR gate error, are introduced accordingly.

\begin{figure}
    \centering
    \includegraphics[width=0.9\linewidth]{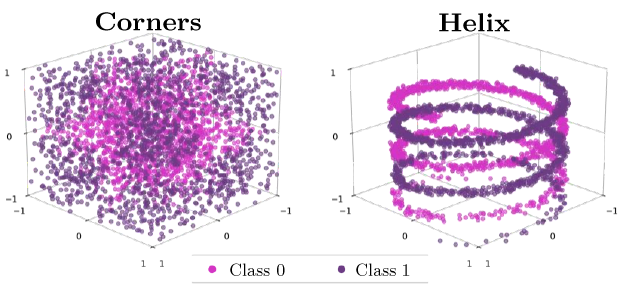}
    \caption{Synthetic datasets created for the numerical evaluation. Left: A 3D cube dataset where points located near the corners belong to one class, and the inner points belong to another. Right: A double helix dataset consisting of two concentric helices with a phase shift of $\pi/2$ between them.}
    \label{fig:datasets}
\end{figure}

\section{Dataset Specifications and Empirical Results} \label{app:datasets_and_results}

In this section, we detail the configuration of our classification tasks and present comprehensive tabular results for our two core experimental evaluations: the impact of re-uploading layers (\textit{Layer Experiment}, Fig.~\ref{fig:results_layers}) and the resilience of the models to varied depolarizing noise (\textit{Noise Experiment}, Fig.~\ref{fig:results_noise}).

\subsection{Experimental Configurations and Benchmarks}

We evaluate our proposed pulse model against its gate-based counterpart across both real-world and synthetic domains. 

\begin{itemize}
    \item \textbf{MNIST (0 vs. 8):} A binary restriction of the MNIST dataset~\cite{mnist_digits} using digits \texttt{0} and \texttt{8}. Samples are flattened into $784$-dimensional vectors and normalized to $[0, 1]$.
    \item \textbf{Iris (Versicolor vs. Virginica):} Sourced from~\cite{iris_53}, using only the two non-linearly separable classes. Features are $4$-dimensional and standardized to zero mean and unit variance.
    \item \textbf{Corners:} A synthetic $3$-dimensional dataset ($d=3$) where Class 0 represents a localized central core near the origin, and Class 1 comprises points distributed towards the outer vertices of a bounded $[-1, 1]^3$ cube (Fig.~\ref{fig:datasets}, Left).
    \item \textbf{Helix:} A synthetic $3$-dimensional dataset ($d=3$) consisting of two intertwined helical strands generated with a relative phase shift of $\pi/2$ (Fig.~\ref{fig:datasets}, Right).
\end{itemize}

\subsection{Experiment Layers: Numerical Results}

This experiment evaluates the model's generalization and expressivity as a function of the architectural depth, systematically varying the number of variational re-uploading layers while keeping the hardware noise profile constant at the baseline level. The structural configurations are determined by the qubit-count and the number of layers, evaluated across multiple random initialization seeds to ensure statistical validity. The primary dependent metrics tracking performance are the final training and test accuracies. The complete numerical breakdown across all evaluated datasets is presented in Table~\ref{tab:results_layers}.

\begin{table*}[ht]
\centering
\small
\setlength{\tabcolsep}{12pt}
\begin{tabular}{cccccccc}
\hline
 & & & \multicolumn{2}{c}{\textbf{Gate-Based Model}} & & \multicolumn{2}{c}{\textbf{Proposed Pulsed Model}} \\ \cline{4-5} \cline{7-8} 
\textbf{Dataset} & \textbf{Qubits} & \textbf{Layers} & $\mathbf{\text{Acc}_{\text{train}}}$ & $\mathbf{\text{Acc}_{\text{test}}}$ & & $\mathbf{\text{Acc}_{\text{train}}}$ & $\mathbf{\text{Acc}_{\text{test}}}$ \\ \hline
Corners & 2 & 1 & $0.694 \pm 0.032$ & $0.691 \pm 0.058$ & & $0.603 \pm 0.049$ & $0.583 \pm 0.042$ \\
 & 2 & 5 & $0.703 \pm 0.103$ & $0.651 \pm 0.079$ & & $0.744 \pm 0.106$ & $0.759 \pm 0.054$ \\
 & 2 & 10 & $0.765 \pm 0.112$ & $0.621 \pm 0.031$ & & $0.813 \pm 0.123$ & $0.743 \pm 0.137$ \\
 & 2 & 15 & $0.779 \pm 0.160$ & $0.569 \pm 0.015$ & & $0.815 \pm 0.122$ & $0.779 \pm 0.118$ \\
 & 2 & 20 & $0.817 \pm 0.127$ & $0.599 \pm 0.047$ & & $0.810 \pm 0.140$ & $0.718 \pm 0.041$ \\
 & 2 & 25 & $0.831 \pm 0.163$ & $0.556 \pm 0.043$ & & $0.841 \pm 0.103$ & $0.707 \pm 0.099$ \\
 & 2 & 30 & $0.889 \pm 0.108$ & $0.569 \pm 0.025$ & & $0.832 \pm 0.117$ & $0.690 \pm 0.123$ \\
 & 2 & 35 & $0.861 \pm 0.161$ & $0.588 \pm 0.035$ & & $0.836 \pm 0.128$ & $0.690 \pm 0.106$ \\
 & 2 & 40 & $0.887 \pm 0.126$ & $0.549 \pm 0.035$ & & $0.864 \pm 0.113$ & $0.708 \pm 0.090$ \\
\cline{2-8}
Digits 0-8 & 2 & 1 & $0.611 \pm 0.054$ & $0.570 \pm 0.054$ & & $0.556 \pm 0.037$ & $0.537 \pm 0.045$ \\
 & 2 & 5 & $0.823 \pm 0.102$ & $0.613 \pm 0.071$ & & $0.880 \pm 0.089$ & $0.826 \pm 0.052$ \\
 & 2 & 10 & $0.859 \pm 0.114$ & $0.606 \pm 0.040$ & & $0.884 \pm 0.124$ & $0.804 \pm 0.081$ \\
 & 2 & 15 & $0.894 \pm 0.108$ & $0.608 \pm 0.075$ & & $0.890 \pm 0.129$ & $0.807 \pm 0.121$ \\
 & 2 & 20 & $0.888 \pm 0.115$ & $0.589 \pm 0.085$ & & $0.887 \pm 0.128$ & $0.790 \pm 0.118$ \\
 & 2 & 25 & $0.884 \pm 0.129$ & $0.573 \pm 0.060$ & & $0.895 \pm 0.109$ & $0.780 \pm 0.063$ \\
 & 2 & 30 & $0.906 \pm 0.136$ & $0.555 \pm 0.073$ & & $0.886 \pm 0.127$ & $0.766 \pm 0.060$ \\
 & 2 & 35 & $0.880 \pm 0.118$ & $0.602 \pm 0.038$ & & $0.806 \pm 0.164$ & $0.730 \pm 0.106$ \\
 & 2 & 40 & $0.913 \pm 0.110$ & $0.580 \pm 0.058$ & & $0.878 \pm 0.090$ & $0.753 \pm 0.095$ \\
\cline{2-8}
Helix & 2 & 1 & $0.617 \pm 0.089$ & $0.583 \pm 0.062$ & & $0.568 \pm 0.042$ & $0.545 \pm 0.045$ \\
 & 2 & 5 & $0.693 \pm 0.162$ & $0.668 \pm 0.071$ & & $0.702 \pm 0.057$ & $0.680 \pm 0.110$ \\
 & 2 & 10 & $0.806 \pm 0.111$ & $0.693 \pm 0.132$ & & $0.867 \pm 0.123$ & $0.823 \pm 0.120$ \\
 & 2 & 15 & $0.852 \pm 0.114$ & $0.700 \pm 0.082$ & & $0.923 \pm 0.126$ & $0.898 \pm 0.042$ \\
 & 2 & 20 & $0.896 \pm 0.106$ & $0.674 \pm 0.091$ & & $0.941 \pm 0.097$ & $0.831 \pm 0.121$ \\
 & 2 & 25 & $0.893 \pm 0.116$ & $0.665 \pm 0.081$ & & $0.925 \pm 0.151$ & $0.831 \pm 0.147$ \\
 & 2 & 30 & $0.892 \pm 0.119$ & $0.652 \pm 0.057$ & & $0.947 \pm 0.078$ & $0.859 \pm 0.114$ \\
 & 2 & 35 & $0.889 \pm 0.138$ & $0.668 \pm 0.082$ & & $0.893 \pm 0.205$ & $0.800 \pm 0.151$ \\
 & 2 & 40 & $0.901 \pm 0.116$ & $0.636 \pm 0.073$ & & $0.916 \pm 0.163$ & $0.834 \pm 0.127$ \\
\cline{2-8}
Iris & 2 & 1 & $0.569 \pm 0.048$ & $0.587 \pm 0.038$ & & $0.600 \pm 0.067$ & $0.587 \pm 0.104$ \\
 & 2 & 5 & $0.797 \pm 0.165$ & $0.560 \pm 0.028$ & & $0.729 \pm 0.154$ & $0.707 \pm 0.166$ \\
 & 2 & 10 & $0.826 \pm 0.118$ & $0.573 \pm 0.043$ & & $0.791 \pm 0.166$ & $0.613 \pm 0.102$ \\
 & 2 & 15 & $0.805 \pm 0.122$ & $0.572 \pm 0.068$ & & $0.783 \pm 0.180$ & $0.656 \pm 0.131$ \\
 & 2 & 20 & $0.882 \pm 0.072$ & $0.567 \pm 0.051$ & & $0.863 \pm 0.147$ & $0.614 \pm 0.102$ \\
 & 2 & 25 & $0.853 \pm 0.103$ & $0.543 \pm 0.016$ & & $0.880 \pm 0.109$ & $0.586 \pm 0.105$ \\
 & 2 & 30 & $0.862 \pm 0.106$ & $0.559 \pm 0.015$ & & $0.887 \pm 0.140$ & $0.626 \pm 0.106$ \\
 & 2 & 35 & $0.853 \pm 0.127$ & $0.562 \pm 0.059$ & & $0.867 \pm 0.076$ & $0.600 \pm 0.109$ \\
 & 2 & 40 & $0.794 \pm 0.113$ & $0.553 \pm 0.036$ & & $0.803 \pm 0.136$ & $0.640 \pm 0.098$ \\
\hline
\end{tabular}
\caption{Comparative performance analysis between the conventional gate-based model and the proposed pulsed model across varying data distributions and layer counts. Statistical metrics denote the $\text{mean} \pm \text{standard deviation}$ calculated over all evaluation trials.}
\label{tab:results_layers}
\end{table*}

\subsection{Experiment Noise: Numerical Results}

For this evaluation, the number of re-uploading layers is strictly fixed at $L = 20$, while the background hardware noise coefficients are locked to the baseline device parameters. To assess model robustness, the independent error probability ($p$) of the depolarizing noise channel is systematically varied across the range $[0.0, 0.3]$. The empirical results, compiled across all evaluation seeds, are summarized in Table~\ref{tab:results_noise}. This setup highlights the performance degradation profile of both the conventional gate-based framework and our proposed pulsed model under escalating environmental decoherence.

\begin{table*}[ht]
\centering
\small
\setlength{\tabcolsep}{12pt}
\begin{tabular}{cccccccc}
\hline
 & & & \multicolumn{2}{c}{\textbf{Gate-Based Model}} & & \multicolumn{2}{c}{\textbf{Proposed Pulsed Model}} \\ \cline{4-5} \cline{7-8} 
\textbf{Dataset} & \textbf{Qubits} & \textbf{Noise} & $\mathbf{\text{Acc}_{\text{train}}}$ & $\mathbf{\text{Acc}_{\text{test}}}$ & & $\mathbf{\text{Acc}_{\text{train}}}$ & $\mathbf{\text{Acc}_{\text{test}}}$ \\ \hline
Corners & 2 & 0.00 & $0.828 \pm 0.107$ & $0.594 \pm 0.036$ & & $0.827 \pm 0.112$ & $0.742 \pm 0.114$ \\
 & 2 & 0.03 & $0.707 \pm 0.089$ & $0.602 \pm 0.046$ & & $0.788 \pm 0.096$ & $0.734 \pm 0.092$ \\
 & 2 & 0.06 & $0.624 \pm 0.075$ & $0.652 \pm 0.064$ & & $0.684 \pm 0.054$ & $0.668 \pm 0.085$ \\
 & 2 & 0.09 & $0.594 \pm 0.069$ & $0.576 \pm 0.072$ & & $0.592 \pm 0.060$ & $0.574 \pm 0.077$ \\
 & 2 & 0.12 & $0.573 \pm 0.084$ & $0.604 \pm 0.067$ & & $0.561 \pm 0.035$ & $0.564 \pm 0.052$ \\
 & 2 & 0.15 & $0.547 \pm 0.059$ & $0.552 \pm 0.063$ & & $0.528 \pm 0.043$ & $0.560 \pm 0.061$ \\
 & 2 & 0.18 & $0.533 \pm 0.035$ & $0.538 \pm 0.055$ & & $0.531 \pm 0.046$ & $0.556 \pm 0.052$ \\
 & 2 & 0.21 & $0.537 \pm 0.051$ & $0.560 \pm 0.090$ & & $0.541 \pm 0.059$ & $0.550 \pm 0.060$ \\
 & 2 & 0.24 & $0.545 \pm 0.067$ & $0.570 \pm 0.092$ & & $0.543 \pm 0.063$ & $0.558 \pm 0.076$ \\
 & 2 & 0.27 & $0.513 \pm 0.014$ & $0.544 \pm 0.044$ & & $0.545 \pm 0.066$ & $0.562 \pm 0.084$ \\
\cline{2-8}
Digits 0-8 & 2 & 0.00 & $0.893 \pm 0.119$ & $0.596 \pm 0.037$ & & $0.890 \pm 0.106$ & $0.792 \pm 0.084$ \\
 & 2 & 0.03 & $0.790 \pm 0.111$ & $0.576 \pm 0.061$ & & $0.859 \pm 0.104$ & $0.787 \pm 0.092$ \\
 & 2 & 0.06 & $0.736 \pm 0.097$ & $0.633 \pm 0.074$ & & $0.852 \pm 0.050$ & $0.797 \pm 0.040$ \\
 & 2 & 0.09 & $0.691 \pm 0.105$ & $0.601 \pm 0.064$ & & $0.838 \pm 0.053$ & $0.804 \pm 0.041$ \\
 & 2 & 0.12 & $0.706 \pm 0.117$ & $0.595 \pm 0.059$ & & $0.835 \pm 0.057$ & $0.773 \pm 0.060$ \\
 & 2 & 0.15 & $0.629 \pm 0.075$ & $0.577 \pm 0.068$ & & $0.821 \pm 0.049$ & $0.737 \pm 0.030$ \\
 & 2 & 0.18 & $0.573 \pm 0.084$ & $0.568 \pm 0.029$ & & $0.751 \pm 0.082$ & $0.681 \pm 0.074$ \\
 & 2 & 0.21 & $0.537 \pm 0.028$ & $0.550 \pm 0.022$ & & $0.620 \pm 0.089$ & $0.621 \pm 0.086$ \\
 & 2 & 0.24 & $0.524 \pm 0.011$ & $0.547 \pm 0.015$ & & $0.554 \pm 0.071$ & $0.568 \pm 0.031$ \\
 & 2 & 0.27 & $0.525 \pm 0.012$ & $0.546 \pm 0.016$ & & $0.525 \pm 0.012$ & $0.546 \pm 0.016$ \\
\cline{2-8}
Helix & 2 & 0.00 & $0.888 \pm 0.122$ & $0.674 \pm 0.052$ & & $0.923 \pm 0.120$ & $0.862 \pm 0.088$ \\
 & 2 & 0.03 & $0.751 \pm 0.124$ & $0.650 \pm 0.062$ & & $0.905 \pm 0.094$ & $0.804 \pm 0.119$ \\
 & 2 & 0.06 & $0.581 \pm 0.046$ & $0.562 \pm 0.029$ & & $0.724 \pm 0.097$ & $0.696 \pm 0.131$ \\
 & 2 & 0.09 & $0.571 \pm 0.055$ & $0.578 \pm 0.052$ & & $0.615 \pm 0.026$ & $0.588 \pm 0.050$ \\
 & 2 & 0.12 & $0.543 \pm 0.055$ & $0.580 \pm 0.085$ & & $0.565 \pm 0.041$ & $0.564 \pm 0.063$ \\
 & 2 & 0.15 & $0.541 \pm 0.056$ & $0.556 \pm 0.059$ & & $0.567 \pm 0.041$ & $0.552 \pm 0.055$ \\
 & 2 & 0.18 & $0.500 \pm 0.000$ & $0.516 \pm 0.034$ & & $0.551 \pm 0.032$ & $0.552 \pm 0.043$ \\
 & 2 & 0.21 & $0.517 \pm 0.035$ & $0.522 \pm 0.046$ & & $0.532 \pm 0.042$ & $0.546 \pm 0.062$ \\
 & 2 & 0.24 & $0.500 \pm 0.000$ & $0.516 \pm 0.034$ & & $0.530 \pm 0.063$ & $0.532 \pm 0.067$ \\
 & 2 & 0.27 & $0.500 \pm 0.000$ & $0.500 \pm 0.000$ & & $0.500 \pm 0.000$ & $0.500 \pm 0.000$ \\
\cline{2-8}
Iris & 2 & 0.00 & $0.883 \pm 0.117$ & $0.549 \pm 0.034$ & & $0.882 \pm 0.103$ & $0.641 \pm 0.085$ \\
 & 2 & 0.03 & $0.828 \pm 0.133$ & $0.563 \pm 0.050$ & & $0.844 \pm 0.138$ & $0.693 \pm 0.145$ \\
 & 2 & 0.06 & $0.700 \pm 0.106$ & $0.584 \pm 0.052$ & & $0.785 \pm 0.124$ & $0.628 \pm 0.122$ \\
 & 2 & 0.09 & $0.640 \pm 0.109$ & $0.597 \pm 0.054$ & & $0.735 \pm 0.154$ & $0.605 \pm 0.099$ \\
 & 2 & 0.12 & $0.576 \pm 0.060$ & $0.581 \pm 0.052$ & & $0.740 \pm 0.146$ & $0.638 \pm 0.118$ \\
 & 2 & 0.15 & $0.558 \pm 0.055$ & $0.579 \pm 0.057$ & & $0.700 \pm 0.168$ & $0.628 \pm 0.116$ \\
 & 2 & 0.18 & $0.559 \pm 0.059$ & $0.571 \pm 0.049$ & & $0.676 \pm 0.167$ & $0.636 \pm 0.117$ \\
 & 2 & 0.21 & $0.556 \pm 0.052$ & $0.562 \pm 0.052$ & & $0.645 \pm 0.126$ & $0.630 \pm 0.114$ \\
 & 2 & 0.24 & $0.555 \pm 0.052$ & $0.559 \pm 0.051$ & & $0.574 \pm 0.063$ & $0.584 \pm 0.071$ \\
 & 2 & 0.27 & $0.552 \pm 0.052$ & $0.557 \pm 0.053$ & & $0.552 \pm 0.052$ & $0.557 \pm 0.053$ \\
\hline
\end{tabular}
\caption{Model robustness and evaluation accuracy under varying levels of depolarizing channel noise ($p$) with depth fixed at $L=20$ layers. Statistical variations reflect the $\text{mean} \pm \text{standard deviation}$ calculated across all independent execution seeds.}
\label{tab:results_noise}
\end{table*}

\pagebreak
\newpage
\textcolor{white}{,}
\vspace{10cm}
\textcolor{white}{,}
\bibliography{main}

\end{document}